# Diffusion kinetic parameters from bulk diffusion limited gas release processes

Ricardo E. Avila

Departamento de Materiales Nucleares, Comisión Chilena de Energía Nuclear,
Cas. 188-D, Santiago, Chile. (ravila@cchen.cl)

The diffusion of a bulk absorbed gas species out of spherical pebbles is studied analytically, stressing the usefulness of the time integral of the diffusion coefficient for analysis of arbitrary temperature schedule experiments. Highly accurate approximations are introduced where the numeric evaluation of the analytic expressions takes considerable time. A method is proposed to extract the diffusion kinetic parameters from a single linear heating ramp, namely, the activation energy of the diffusion coefficient and the ratio of the corresponding preexponential factor to the radius of spherical pebbles.

The release of a gas from a material in round pebble form is of interest towards the generation of tritium by neutron bombardment of lithium ceramics. The release kinetics from several oxides has been studied and interpreted in terms of surface desorption control[1]. However, a discussion remains regarding the possibility of the release being controlled by the bulk diffusion[2,3].

As a tool for these studies or for other diffusion controlled processes, in the form of a gas, or into a liquid, the diffusion equation is solved, in this work, for an arbitrary temperature schedule. Numerical methods are introduced to extract the process kinetics from isolated peaks in a single linear temperature ramp. The method is exemplified by application to somewhat realistic synthetic data. For completeness, the solution of the diffusion equation will be outlined, next, with emphasis on its time dependence.

The equation for diffusion of a gas out of spherical pebbles can be written as

$$\nabla^2 u(r,t) = \frac{1}{k_o \kappa(t)} \frac{\partial u}{\partial t}, \quad (1)$$

where $u(r,t)$ is the bulk gas concentration, $r$ is the radius, $t$ is the time, $k_o \kappa(T)$ is the diffusion coefficient, with $k_o$ carrying the dimensions of length squared per unit time, and $\kappa(t)$ being the adimensional temperature activated factor

$$\kappa(t) = \exp(-E_d/(k_B T(t))). \quad (2)$$

Here, $E_d$ is the diffusion activation energy, $k_B$ is the Boltzmann constant, and $T$ is the absolute temperature.

Assuming spherically symmetric boundary conditions, Eq. (1) reduces to an equation involving the radius, $r$, and the time, $t$, only, which can be solved by separation of variables. Writing $u(r,t) = X(r)Y(t)$, where $X(r) = x(r)/r$, and demanding that $X$ be finite at the origin leads to $x(r)$ in terms of $sin(cr)$, where $-c^2$ is the separation constant.

The equation for $Y(t)$

$$\frac{dY}{dt} = -c^2 k_o \kappa(t) Y \quad (3)$$

can be turned into a simple equation with constant coefficients with the substitution[4]

$$\Lambda(t) = \int_0^t \kappa(t)\,dt, \quad (4)$$

which is a monotonically increasing function of time.

Introducing the physical assumptions, the case of bulk limited release (BLR) is considered, where the hindrance to the gas release is the diffusion to the surface; from which the gas desorbs readily. The latter condition is expressed, for a sphere of radius $a$, by the condition $u(a,t) = 0$, from which the separation constant $-c^2$ follows.

Then, assuming a uniform gas concentration, $u_o$, at $t = 0$, throughout the sphere, determines the coefficients in the series expansion of $u$ as

$$u(r,\lambda) = \frac{2au_o}{\pi r} \sum_{n=1}^{\infty} \frac{(-1)^{n+1}}{n} \sin\left(\frac{n\pi r}{a}\right) \exp\left[-n^2 \pi^2 \lambda\right] \quad (5)$$
,

where the adimensional release coefficient, $\lambda(t) = (k_o/a^2)\Lambda(t)$.

The total gas released up to a certain value of the release coefficient, $\lambda$, is obtained as the initial charge, $U_o = 4\pi a^3/3$, minus the integral of Eq. (5) over the sphere. The resulting expression, normalized to $U_o$, is

$$c(\lambda) = 1 - \frac{6}{\pi^2} \sum_{n=1}^{\infty} \frac{1}{n^2} \exp\left[-n^2\pi^2\lambda\right]. \quad (6)$$

Eq. (6) is well known for isothermal processes, and the extension to arbitrary temperature schedules, by the use of the $\lambda$ parameterization, is a direct consequence of the transformation in Eq. (4). Next, a useful approximation is introduced, and a method to extract the diffusion kinetics is proposed.

The expression (6) can be approximated for short times, as

$$c(\lambda) \cong 6\sqrt{\lambda/\pi} - 3\lambda. \quad (7)$$

This approximation matches $c(\lambda)$ within the machine precision used for this work (12 significant digits) up to the point where $\lambda \sim 0.05$, corresponding to $c(\lambda) \sim 0.6$. From that point on, the deviation increases rapidly, reaching 0.01% when $c(\lambda) \sim 0.84$, and 1% when $c(\lambda) \sim 0.96$. Clearly, these values depend on the system parameters ($k_o$, $a$, $E_d$), and on the specific temperature schedule $T(t)$ only through the combined $\lambda$ parameter. As a guide, it will be noted, below, that over a broad range of the system parameters, a linear temperature ramp leads to $c(\lambda) \sim 0.66$ by the release maximum.

The derivative

$$\frac{d\ln c(\lambda)}{d\lambda} \xrightarrow{\lambda \to 0} \frac{1}{2} \quad (8)$$

and it deviates down, by 1% by $c(\lambda) \sim 0.055$, and by 5% around $c(\lambda) \sim 0.302$. In contrast, it can be shown that in surface desorption limited gas release the same expression approaches a value of 1. This contrast has been pointed out as a means of distinguishing bulk limited from surface desorption limited tritium release from ceramic pellets in isothermal experiments. Expressing the released charge in terms of $\lambda$ provides an extension of that observation to arbitrary temperature schedules. Thus, towards the use of Eq. (8) in an isothermal release experiment, the initial temperature rise does not represent a limitation or source of error, as long as $\lambda$ is consistently evaluated along the temperature transient and into the predetermined isothermal stage.

The normalized bulk diffusion limited gas release rate is obtained by differentiation of Eq. (6)

$$R(\lambda) = 6(k_o/a^2)\kappa(t(\lambda))\sum_{n=1}^{\infty} \exp\left[-n^2\pi^2\lambda\right]. \quad (9)$$

Where, to an accuracy of $10^{-6}$,

$$\sum_{n=1}^{\infty} \exp(-n^2 x) = \begin{cases} \frac{1}{2}\left[\vartheta_3(0, e^{-x}) - 1\right]; & x < 3.7 \\ \exp(-x); & x > 3.7 \end{cases}. \quad (10)$$

where $\vartheta_3(u, q)$ is the elliptic theta function of the third kind[5] of which numerical implementations exist. In Eq. (10), the top expression may be used over most of the release peak, where $x$ is in the range of $10^{-3}$ to 10. At the high temperature end of the peak, with $x > 10$, the top expression, may run into round off errors. If a numerical implementation of the theta function is not available, a few tens and up to several thousand terms are required for convergence of the series at the low temperature side of the release peak.

Alternatively, a highly accurate approximation follows from Eq. (7),

$$R(\lambda) \cong 3(k_o/a^2)\kappa(t(\lambda))\left[1/\sqrt{\pi\lambda} - 1\right], \quad (11)$$

of which, over a linear temperature ramp, the deviation from Eq. (9) is $10^{-4}$ % at the peak, and 0.1% (1%), when $R$ has dropped past the release peak by 10% (22%) of the maximum value.

The time, $t_p$, at which a release maximum occurs can be obtained from Eq. (9). However, the peak of a linear ramp, $T(t) = T_o + \beta t$, is well within the applicability range of the approximation (11), from which the simpler expression

$$\left[1 - (\pi\lambda(t_p))^{1/2}\right]\frac{\beta E_d}{k_B T_p^2} \cong \frac{k_o \kappa(T_p)}{2a^2 \lambda(t_p)}, \quad (12)$$

results, where $T_p = T(t_p)$, and $\lambda$ should not be written in terms of $T$, since $\lambda(T)$ is not necessarily single-valued. Eq. (12) is valid, also, for local maxima of arbitrary temperature schedules, with $\beta = dT/dt > 0$, and $d\beta/dt = 0$.

The expression (9), is valid for arbitrary (analytical or experimental) temperature profiles. By packing the time dependence of the diffusion coefficient into $\lambda(t)$ it is not necessary to solve the differential equations for each $T(t)$ schedule. Furthermore, given the easy of modern computerized data acquisition, Eq. (4) can be evaluated along an arbitrary experimental temperature schedule, and the result fed to Eqs. (5) or (9) to fit model parameters or other analysis. Such fitting without the use of the $\lambda(t)$ parameterization would, in general, imply the numeric solution of the corresponding differential equation for each combination of the trial values of the parameters. In comparison, the use of Eq. (4) in the otherwise analytical expressions requires a much simpler numerical integration.

Towards the use of the methods given above, the time integral of the diffusion coefficient is necessary. This is given, next, for the two most used temperature schedules: the isothermal and linear ramp processes.

In the isothermal process, $T_1(t) = T_o$, and $\Lambda_1(t) = \kappa(T_o) t$.

For linear ramps, $T_2(t) = T_o + \beta t$,

$$\Lambda_2(T) = \frac{1}{\beta}[L_2(T) - L_2(T_o)], \quad (13)$$

where

$$L_2(T) = \int \exp\left(\frac{-E_d}{k_B T}\right) dT = T \exp\left(\frac{-E_d}{k_B T}\right) + \frac{E_d}{k_B} \mathrm{Ei}\left(\frac{-E_d}{k_B T}\right) \quad (14)$$

The numerical evaluation of the exponential integral, Ei(x),

$$\mathrm{Ei}(T) = -\int_{-T}^{\infty} \frac{\exp(-u)}{u} du, \quad (15)$$

where the principal value is taken, has been studied by Chen[6], who suggests to take up to the smallest term ($k \sim x$) in the asymptotic expansion (Ref. [5], 8.214.3),

$$\mathrm{Ei}(-x) = e^{-x} \sum_1^n (-1)^k \frac{(k-1)!}{x^k}, \quad (16)$$

and add one half of the next term.

For most applications, where essentially no release occurs at $T_o$, $L_2(T_o)$ can safely be ignored in Eq.(13).

The release rate and charge curves shown in Fig. 1 have been calculated using the expressions and approximations given above. The release rate, normalized to its peak value, is calculated with $k_o/a^2 = 5 \cdot 10^4$ s$^{-1}$, $\beta = 5$ °C/min., and $E_d = 1.219$ eV (28.1 kcal/mole), which is chosen to set the peak position at 450 °C.

The peak temperature Eq.(12) allows for exploring the behavior of $T_p$ against variations in the system parameters. Thus, while keeping other parameters as written above, varying the activation energy over a wide range (0.7 to 4 eV), the peak temperature rises close to linearly from 153 to 1976 °C, and the peak occurs over a narrow range, 66.59 to 66.97%, of the released charge. Varying the heating rate from 0.01 to 10 °C/min. moves the peak temperature from 287 to 474 °C, while the released charge diminishes from 67.82, to 66.56%. Over this heating rate range, the slope of $\ln(\beta)$ vs. $1/T_p$ is nearly constant. The evaluation of this slope, as the corresponding implicit derivative, using Eq. (12), contains several terms of similar value, involving the system parameters, and the values of $T_p$, $\kappa(T_p)$ and $\lambda(T_p)$, from which no clear relation to the system parameters has been noted. However, if other methods should fail, that expression could be solved for the activation energy, if a few $(\beta, T_p(\beta))$ pairs can be determined, and a reasonable value of $k_o/a^2$ can be estimated.

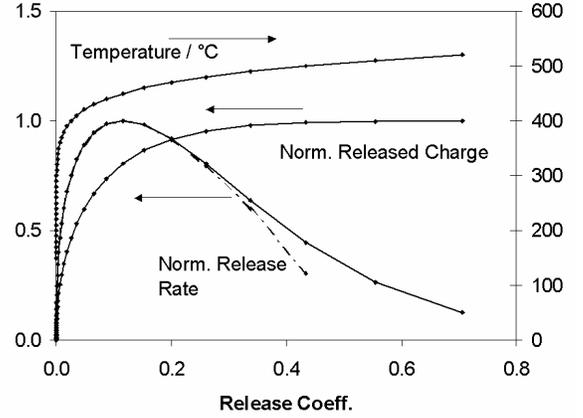

Fig. 1. Bulk limited gas release from spheres of radius 1 mm, with $k_o = 500$ cm$^2$/s, and $E_d = 1.219$ eV. Using the coefficient $\lambda$ as abscissa makes the released charge curve applicable to all systems. The bottom dotted line is the numerical approximation

A procedure has been devised for analyzing BLR data. Given a single-peak BLR curve, the function $c(\lambda)$ can be identified as

$$c(\lambda(t)) = 1 - U(t)/U_o, \quad (17)$$

where

$$U(t) = \int_t^{t_f} R(t')dt', \quad (18)$$

$t_f$ is the time at the high-temperature end of the experiment, and $U_o = U(t=0)$.

Then, Eq. (6), or, over most of the release peak, the approximation (7), can be solved, numerically, for $\lambda(t)$, of which the time derivative is interpreted as $k_o\kappa(t)/a^2$. Finally, the Arrhenius analysis of the latter expression provides $-E_d/k_B$ as the slope, and $\ln(k_o/a^2)$ as the $1/T \to 0$ intercept. The numeric implementation of this procedure has been tested using the logarithm of the $c(t)$ and $\lambda(t)$ curves.

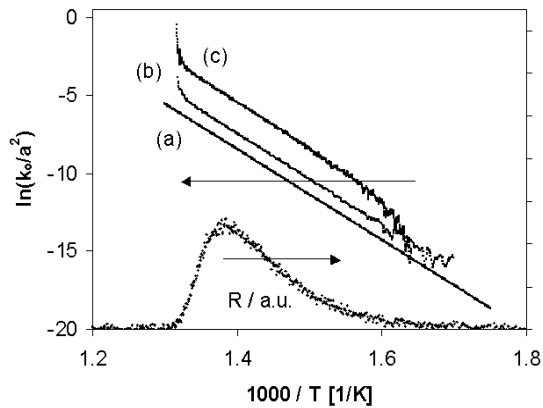

Fig. 2. Arrhenius analysis of noisy synthetic BLR data. Curve (a): from exact calculation, (b): from Gaussian noise, 2.5% of value + 2.5% of peak value added, (c): from BLR curve truncated at 1% of peak maximum, plus same noise as in (b). Curves (b) and (c) shifted upwards by 1 and 2, respectively, for clarity. Curve R: noisy BLR leading to analysis in (b).

This approach is sensitive to noise, and to errors in the determination of the initial gas charge, $U_o$, which lead to artifacts at both the low and high ends of the temperature range. These sensitivities have been tested, as shown in Fig. 2, by introducing errors on synthetic data calculated for spherical pebbles, with $k_o/a^2 = 5 \cdot 10^{14}$ s$^{-1}$, $\beta = 5$ °C/min., and $E_d = 2.505$ eV (57.73 kcal/mole), which is chosen to set the peak position at 450 °C. The deviation of the calculated $k_o/a^2$, and $E_d$ from the input values has been controlled by restricting the final Arrhenius analysis to the central region of the $k_o\kappa(t)/a^2$ vs. $1/T$ curve, corresponding to $c(\lambda)$ from approximately 0.1 to 0.9.

The addition of normally distributed noise (curve (b)), at a level of 2.5% of the peak value, plus 2.5% of "measurement", leads to deviations of $k_o/a^2$, and of $E_d$ of the order of 1 and 2%, respectively. Then, in addition to the previous noise, both the initial and final times for the integration in Eq. (18) have been taken at a release rate of 5% of the maximum value. In this case, a 1.5 % error appears in $U_o$, leading to 2, and 3% error in $E_d$, and $k_o/a^2$, respectively.

The derivative $d\ln(\lambda)/dT$ is particularly sensitive to noise. This can be controlled, at each experimental time, by fitting a quadratic polynomial to several points (more than 3) around that specific time, and calculating the derivative from the polynomial fit. The resulting derivative is seen to be highly noise immune if 9 points are used for the fit, over a curve in which the peak is spread over 120 points at half height.

The present method can be extended to systems other than the initially uniformly charged spherical pebbles, considered here, as long as the released charge curve can be calculated as a function of $k_o\Lambda(t)$.

In summary, a recollection of the theory of diffusion out of spherical pebbles has been presented, leading to the analytical analysis of experimental data following arbitrary temperature schedules. Highly accurate approximations have been introduced facilitating the numerical calculations, and providing additional insight on the physical behavior over most of the release peak. A method is exemplified to extract the diffusion kinetics from a single linear heating ramp experiment.

## References

[1] C. Alvani, P.L. Carconi, St. Casadio, and S. Casadio, Proceedings of *CBB-10* (Tenth International Workshop on Ceramic Breeder Blanket Interaction), 22.-24. October, Karlsruhe, Germany, 2001.
[2] P. Bertone, J. Nucl. Materials, **151**, 281 (1988).
[3] T. Tanifuji, D. Yamaki, S. Nasu and K. Nodfa: J. Nucl. Mater. 238–2563 (1998) 543.
[4] H. S. Carslaw and J. C. Jaeger: *Conduction of Heat in Solids* (Oxford University Press, N. York, USA, 1959) Sec. 1.6, eq. (10), and footnote on p. 11.
[5] I.S. Gradshteyn and I.M. Ryzhik, *Table of Integrals, Series, and Products,* Sec. 8.18, Academic Press, Inc., (1980).
[6] R. Chen, J. Comput. Phys. **8**, 156, (1971).